\documentclass{elsarticle}
\usepackage{amsmath,amsthm,latexsym,subfig,verbatim}
\newcommand{\be}{\begin{equation}}
\newcommand{\ee}{\end{equation}}
\newcommand{\bea}{\begin{eqnarray}}
\newcommand{\eea}{\end{eqnarray}}
\newcommand{\ba}{\begin{array}}
\newcommand{\ea}{\end{array}}

\begin{document}

\begin{frontmatter}

\title{Algorithm for multivariate data standardization up to third moment}


\author{Vadim Asnin}

\address{Radar Division, Elta Electronics Industries, Ashdod,
Israel\footnote{Email: vasnin@elta.co.il}}

\begin{abstract}
An algorithm for transforming multivariate data to a form with
normalized first, second and third moments is presented.
\end{abstract}

\begin{keyword}

Multivariate data, data standardization, third moment.

\end{keyword}

\end{frontmatter}

\section{Introduction}
\label{Introduction}

Statistical analysis of multivariate data is a classical problem
encountered in essentially every field of research. Because of
importance of this problem numerous approaches have been proposed
over the years. Differences in algorithms come both from special
features of data sets they are intended to be used for and from
specific questions that they are supposed to address. Special
features of data sets may include an approximate Gaussianity (that
may be a case and may be not), a possible presence of different
classes in the data (which is to be recognized), a possibility to
employ training sets or impossibility to do so etc. Purposes of
algorithms can also be very different, like noise reduction,
determination of leading statistical trends, discovery of anomalous
data points etc.

Many of existing algorithms are based on an idea of normalizing a
covariance matrix of the data distribution (Principal Component
Analysis (PCA) \cite{PCA} and its generalizations, e.g.
\cite{NIPALS,CA,WPCA}). Among them is an RX algorithm \cite{RX},
which is widely used for analyzing hyperspectral data. In recent
years there appeared some generalizations of this procedure which
deal with higher moments of data distribution (they are based on CP
\cite{CanDecomp,PARAFAC} and Tucker \cite{Tuck} decompositions and
various generalizations \cite{BW,KB}).

The essential idea of the RX procedure is to normalize a spread of
the data distribution in all directions to be unity. Then it is
legitimate to compare between different directions. It is convenient
to reformulate this as a normalization of the first two moments of
the data distribution. Stated this way, the procedure allows the
following generalization: go to higher moments and normalize them as
well to be those of, say, a multivariate Gaussian distribution. This
problem has been recognized and partially treated (see, for example,
\cite{BW,ZO}; for slightly non-Gaussian data one can use
Gram-Charlier or Edgeworth expansions). In the present paper we
propose an algorithm which completely solves this problem for the
third moment for most of practically relevant cases.

\section{Setup and formulation of the problem}

We consider the following situation. Suppose that there is a given
distribution of data points in $N$-dimensional space. In order to
``standardize" the distribution there exists the following standard
procedure (RX):
\begin{itemize}
\item Compute a mean of the distribution and subtract it. After
this step a mean value of all $x_i$'s is zero.\item Compute a
correlation matrix of the new distribution. Identify its
eigenvectors (they are orthogonal) and choose them to be a basis in
the space. Rotate the distribution to this basis. Normalize the RMS
of each dimension to be unity.\end{itemize} After performing these
steps one obtains a distribution with no correlations between the
modes.

We choose to reformulate the procedure above in the following way.
Suppose that our data points represent samples of a certain
(unknown) PDF. Then the first step above cancels out a first moment
of this PDF, and the second step normalizes a second moment to be a
unity matrix. Then one would like to continue this procedure and to
normalize higher moments as well. Were one able to normalize all
moments of the distribution, he would end up with the multivariate
Gaussian distribution, and along the computation he would eventually
discover a coordinate system in which the underlying PDF is a
Gaussian. In practice one would restrict himself to a finite number
of moments. A normalization of the third moment is a subject of the
present paper.

It should be noted that whence the first moment of a multivariate
distribution is a vector and the second moment is a symmetric
matrix, the third and higher moments are multidimensional tensors
(the third moment, for example, is a three-dimensional tensor). It
is a well-known fact that it is much harder to treat such tensors
than matrices, mainly because a lack of an analog of a
diagonalization procedure, despite an existence of some analogs
(like the CP and Tucker decompositions \cite
{CanDecomp,PARAFAC,Tuck}, see also \cite{GER} for a different kind
of a generalization). In addition, to normalize the first two
moments it is enough to use just linear transformations, but for
higher moments a transformation is necessarily highly nonlinear (one
might expect a transformation for a third moment to be quadratic,
and we will see that this is indeed correct, see also \cite{ZO} for
one-dimensional case). But then the underlying logic of RX is
inapplicable in a following sense: In RX one essentially determines
a small set of points (eigenvectors) which reproduce a second moment
of the distribution, and then carries out a linear transformation
which normalizes it. By linearity, the same transformation will
normalize the second moment of the whole distribution. But for
nonlinear transformations such an argument will not work. In the
next section we describe an algorithm which circumvents both
problems.

\section{Description of algorithm}

Suppose that there is a distribution of data points in
$N$-dimensional space with coordinates $x_i$. We assume that the
first two moments of it have been normalized before (by means of
RX). The distribution still possesses a third moment
$Q_{ijk}=\langle x_i x_i x_k\rangle$, where $i,j,k=1...N$. In order
to remove this moment from the distribution we implement the
following procedure.

All data points are spread in the $N$-dimensional space. Consider an
addition of a single dimension to this space and denote a new
coordinate by $z$. In this extended $N+1$-dimensional space all data
points belong to a subspace with $z=0$, which we denote by $H$ (for
``horizontal").

\begin{figure}[h!]\centering
\subfloat[Data
points]{\label{LiftingFigureInitialData}\includegraphics[height=0.15\textheight]{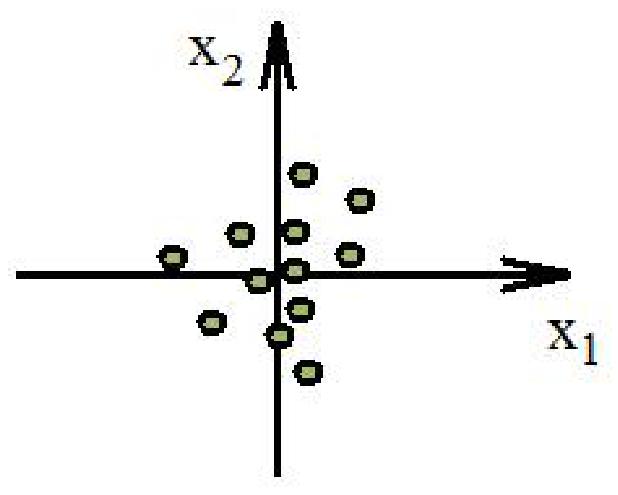}}\quad
\subfloat[Single dimension
added]{\label{LiftingFigureDimensionAdded}\includegraphics[height=0.15\textheight]{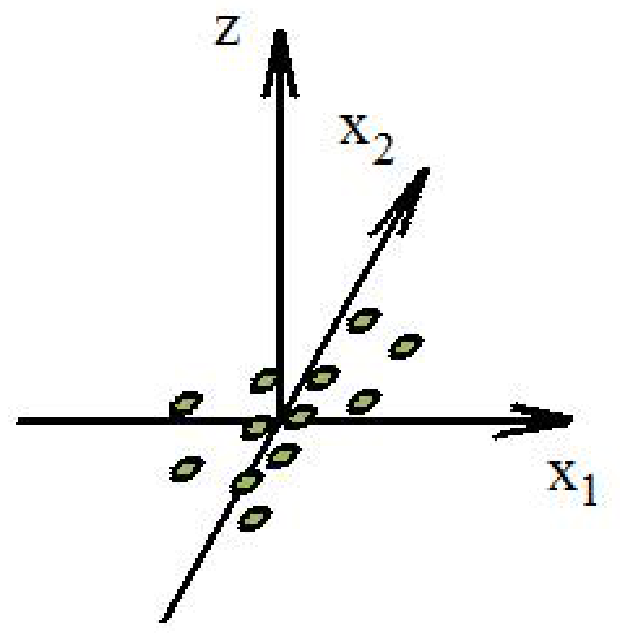}}\quad
\subfloat[Data lifted (only a few points
showed)]{\label{LiftingFigurePointsLifted}\includegraphics[height=0.15\textheight]{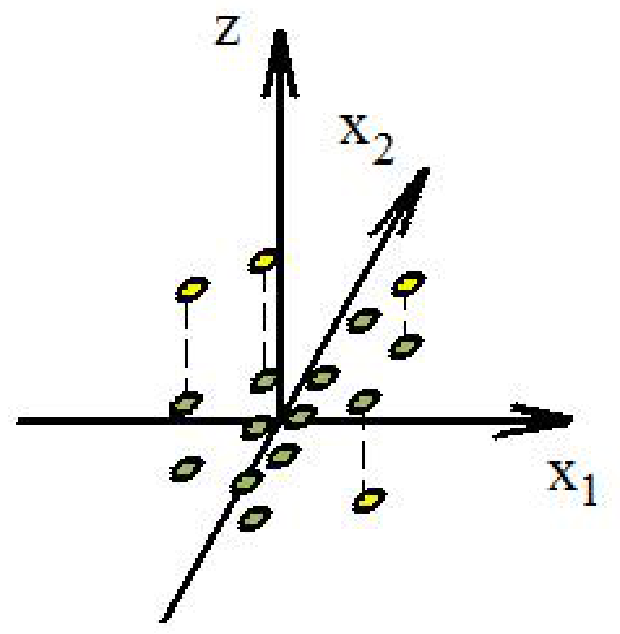}}
\caption{Lifting of data points}\label{LiftingFigure}
\end {figure}

Now we want to ``lift" all data points in the $z$-direction. It
means that we want to assign each data point a certain
$z$-coordinate in a way that the new data distribution would satisfy
certain requirements. We require that the new distribution will
possess trivial first and second moments, or, formally, \be \langle
z\rangle=0,\qquad \langle z\, x_i\rangle=0, \qquad \langle
z^2\rangle=1 \label{LiftEq}\ee In order to satisfy the first two
requirements it is enough to choose (here and throughout the rest of
the paper we use the Einstein convention for repeated indices) \be
z=\alpha + \beta_i\,x_i+\gamma_{ij}\,x_ix_j,\label{Lift}\ee where
the coefficients are given by \be \alpha=-\gamma_{ii}, \qquad
\beta_i=-Q_{ijk}\gamma_{jk}.\label{Coeffs}\ee We see that a lift is
completely defined by a symmetric matrix $\gamma_{ij}$: with this
matrix given one computes the rest of the coefficients from eq.
\ref{Coeffs} and then the $z$-coordinates of data points from eq.
\ref{Lift}. The last requirement of eq. \ref{LiftEq} can be then
satisfied by a change in the overall normalization of the
$z$-coordinate. In fig. \ref{LiftingFigure} we demonstrate the
process of data lifting. In fig. \ref{LiftingFigureInitialData}
there is an initial distribution of data, in fig.
\ref{LiftingFigureDimensionAdded} there is a space with a single
dimension added, and in fig. \ref{LiftingFigurePointsLifted} there
are lifted data points (in order not to abuse the figure we show a
lift of few points only).

Having carried out the lift, we may consider various orthogonal
rotations in the extended space. Such a rotation will change
coordinates of data points and, in particular, their projections
onto the subspace $H$. Orthogonal rotations do not change neither
first nor second moments provided that they have been normalized.
However, the third moment tensor of the lifted distribution will
rotate together with data points and its projection on the subspace
$H$ will change. Therefore we have at this point a tool to change
the third moment of the distribution without destroying the first
two. We would like to choose a rotation of the lifted distribution
in a way that would minimize the third moment of the new
distribution in the subspace $H$. It follows from this construction
that the coordinates of any new data point in $H$ are quadratic
functions of its initial coordinates.

Recall that we still have a freedom to choose a symmetric matrix
$\gamma_{ij}$, which defines a lift. In order to get rid of this
ambiguity we consider an addition of many such $z$-coordinates, one
for every independent component of $\gamma_{ij}$. There are
$N(N+1)/2$ such components, so we add this amount of dimensions and
denote them by $z_{\xi}$. Each dimension is added independently of
others in the way described above. At the end we get a distribution
which satisfies the following conditions, which generalize the first
two of eq. \ref{LiftEq} to higher-dimensional case: $\langle
z_{\xi}\rangle=0, \quad\langle z_{\xi}\,x_i\rangle=0$. In order to
have completed a normalization of the second moment (i.e. to provide
that $\langle z_{\xi}\,z_{\eta}\rangle=\delta_{\xi\eta}$) it remains
to normalize the third moment of the $z$-coordinates by means of RX
in $N(N+1)/2$-dimensional space. With this have been done, we obtain
a distribution in the $N+N(N+1)/2$-dimensional space with normalized
first two moments. Then we want to rotate it in a way that will
cancel a third moment of the projection onto the $N$-dimensional
subspace $z_{\xi}=0$, which we still denote by $H$.

We turn now to a description of a numerical procedure for computing
the necessary rotation. It is by no means necessary to compute it
this way, one can follow a different route instead. For instance,
one can use  some kind of Tucker decomposition of the third order
tensor, following the ideas of \cite{AH,RTB}, or employ a
modification of an Alternating Least Squares algorithm (a commonly
used version of it will produce some sign differences in rotation
matrices of different tensor dimensions, which is unacceptable).

Denote a third moment of the lifted distribution by
$Q_{\mu\nu\lambda}$, where the indices run over the values of
$(i,\xi)$. We endow it with a norm \be
\|n\|^2=Q_{ijk}Q_{ijk}=P_{\mu\mu'}P_{\nu\nu'}P_{\lambda\lambda'}Q_{\mu\nu\lambda}Q_{\mu'\nu'\lambda'},\ee
 where the indices in the first expressions run only over the $z_{\xi}=0$
 subspace $H$, whereas in the second they run over the whole space, but
 there appears a matrix $P$ which carries out a projection onto $H$.

Consider orthogonal rotations $A_{\mu\nu}$ (the orthogonality
conditions that they should satisfy are
$A_{\mu\lambda}A_{\nu\lambda}=\delta_{\mu\nu}$), under which tensors
transform as \be
Q'_{\mu\nu\lambda}=A_{\mu\alpha}A_{\nu\beta}A_{\lambda\gamma}Q_{\alpha\beta\gamma}.\ee
Introduce infinitesimal rotations which mix two types of dimensions.
A matrix of such a rotation can be written in a block form as \be
A=\exp\biggl(\left[\ba{cc} 0 & -\phi \\ \phi^T & 0\ea\right]\biggr)\simeq\left(\ba {cc} 1 & -\phi \\ \phi^T & 1 \\
\ea \right),\ee where upper-left corner is of dimension $N\times N$
and the lower-right corner is of dimension $N(N+1)/2\times
N(N+1)/2$. The projection
matrix $P$ introduced above is of the form \be P=\left( \ba {cc} 1 & 0 \\
0 & 0\ea\right).\ee A norm of the rotated tensor is \be
\|n'\|^2=(A^T\,P\,A)_{\mu\mu'}(A^T\,P\,A)_{\nu\nu'}(A^T\,P\,A)_{\lambda\lambda'}Q_{\mu\nu\lambda}Q_{\mu'\nu'\lambda'},\ee
where an explicit form of the matrix $A^T\,P\,A$ is \be
A^T\,P\,A=\left(\ba{cc} 1&-\phi\\-\phi^T&0\ea\right).\ee Then a
change in the norm under such a rotation to the first order in
$\phi$ is \be \delta\|n\|^2 =-6\,\phi_{i\xi}Q_{ijk}Q_{\xi jk}.\ee We
see that if we choose $\phi$ to be proportional to \be
\Phi_{i\xi}=Q_{ijk}Q_{\xi jk}\label{Phi}\ee then we achieve the
fastest decrease of the norm. Therefore we can carry out a gradient
descent computation, where at every step we choose a rotation matrix
to be of the form \be A=\exp\biggl(\left[\ba{cc} 0
&-\Omega\Phi\\\Omega\Phi^T & 0 \ea\right]\biggr),\ee where $\Omega$
is some small parameter. The full rotation matrix is a
multiplication of matrices obtained at the intermediate steps. Along
this gradient descent flow the following differential equations for
components of $Q_{\mu\nu\lambda}$ hold: \bea \dot
Q_{ijk}=-\Phi_{i\rho}Q_{\rho jk}-\Phi_{j\rho}Q_{i\rho k}-\Phi_{k\rho}Q_{ij\rho}\\
\dot Q_{ij\xi}=-\Phi_{i\rho}Q_{\rho j\xi}-\Phi_{j\rho}Q_{i\rho \xi}+\Phi_{m\xi}Q_{ijm} \\
\dot Q_{i\xi\eta}=-\Phi_{i\rho}Q_{\rho\xi\eta}+\Phi_{m\xi}Q_{im
\eta}+\Phi_{m\eta}Q_{i\xi m} \\ \dot
Q_{\xi\eta\zeta}=\Phi_{m\xi}Q_{m\eta\zeta}+\Phi_{m\eta}Q_{\xi m
\zeta}+\Phi_{m\zeta}Q_{\xi \eta m}\eea with $\Phi_{i\xi}$ defined in
eq. \ref{Phi}. At the final point of the evolution $\Phi_{i\xi}$
will vanish. Since a pair of indices $i,j$ accepts $N(N+1)/2$
values, the same amount as does $\xi$, one can think of the result
$\Phi_{i\xi}\equiv Q_{ijk}Q_{\xi jk}=0$ as a matrix $Q_{\xi jk}$ of
dimension $N(N+1)/2\times N(N+1)/2$ acting on $N$ vectors in
$N(N+1)/2$ - dimensional space (written as $Q_{ijk}$) with vanishing
result. If the matrix $Q_{\xi jk}$ is non-degenerate then it is
possible only if $Q_{ijk}=0$. So the only possibly problematic
points in the space of tensors are those at which $Q_{\xi jk}$ is a
degenerate operator. At these points the first order expansion of
the norm is not sufficient; to the first order there is no change in
the norm and the question is whether such a point is a local minimum
of the norm or it is rather a saddle point or even a local maximum.
In the former case the algorithm will get stuck there (if it reaches
a vicinity of it), whereas in the latter cases it will escape that
point.

An expansion of the norm to the second order in $\phi_{i\xi}$ around
a point with $Q_{ijk}Q_{\xi jk}=0$ is \begin{multline}
\delta\|n\|^2=\\-\phi_{i\xi}\phi_{l\xi}Q_{ijk}Q_{ljk}+\phi_{k\xi}\phi_{k\eta}Q_{\xi
ij}Q_{\eta ij}+2\phi_{i\xi}\phi_{j\eta}Q_{\eta ik}Q_{\xi
jk}+2\phi_{i\xi}\phi_{j\eta}Q_{ijk}Q_{\xi\eta k}\end{multline} In
the last term in this expression there appear components $Q_{\xi\eta
i}$ of the tensor,  components which did not appear at all in the
discussion above. If they are large enough they can in principle
make this expression positively definite. From the construction it
is clear that these components are related to higher moments of the
initial distribution (up to fifth moment). So, if these higher
moments are very large the algorithm may not find a rotation that
would cancel the third moment. In the next section we discuss some
possible approaches to a solution of this problem. However, in all
practical situations considered by the author the algorithm
converged to 0.

\begin{figure}\centering
\subfloat[Initial distribution of data
points]{\label{ExampleFigureInitialData}\includegraphics[height=0.25\textheight]{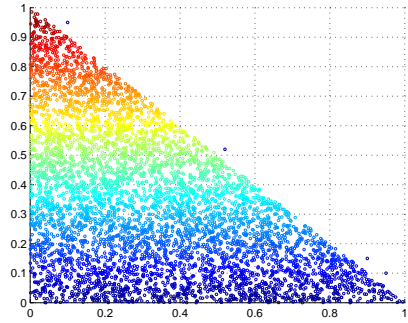}}\quad
\subfloat[Distribution of data points with first and second moments
normalized]{\label{ExampleFigureSecondMoment}\includegraphics[height=0.25\textheight]{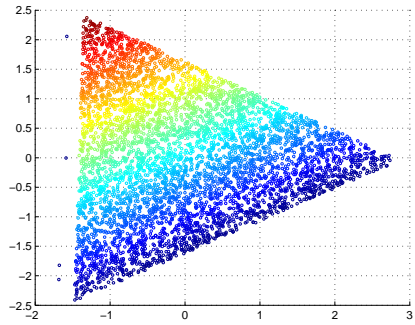}}\quad
\subfloat[Distribution of data points with three moments
normalized]{\label{ExampleFigureThirdMoment}\includegraphics[height=0.25\textheight]{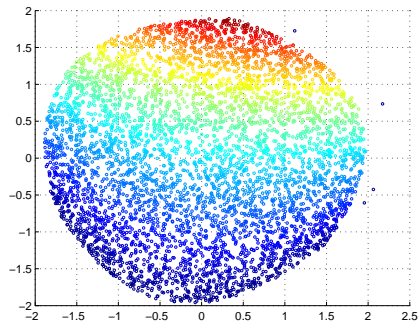}}
\caption{Example of third moment removal}\label{ExampleFigure}
\end{figure}

In fig. \ref{ExampleFigure} we present a result of this computation
on a simulated two-dimensional data. The data points are initially
distributed homogeneously within a triangle $x\geq 0$, $y\geq 0$,
$x+y\leq 1$. In addition, there are four anomalous data points above
the diagonal. A coloring of the points is introduced in order to
clarify a distortion of the distribution when more and more moments
have been normalized. In fig. \ref{ExampleFigureSecondMoment} there
is a distribution with first two moments normalized (in other words,
the result of RX). The triangle has become an equilateral one (which
is more symmetric), but anomalous data points are still not of the
biggest norm. In fig. \ref{ExampleFigureThirdMoment} there is a
distribution with three first moments normalized. The distribution
is almost circular, and the anomalous data points are of the biggest
norms.

\section{Conclusions and summary}
In this paper we described an algorithm for a standardization of the
third moment of data distributions and presented an example of its
operation. The algorithm consists of two steps: \begin{itemize}\item
Lift the data by adding $N(N+1)/2$ dimensions to initial
$N$-dimensional space. Values of new coordinates are quadratic
functions of initial coordinates.
\item Rotate a new distribution so that in the third moment of the
projection onto the horizontal subspace would vanish.
\end{itemize} Some remarks here are in order.

Firstly, the author is neither aware of an analytic solution for a
rotation that would make the third moment of the projection vanish,
nor was he able to solve for it, and therefore had to implement a
numeric solution described above. Moreover, as shown in the previous
section, there might in principle be situations where the algorithm
in its present form will not converge. It is a challenging problem
to find such an analytic solution. A knowledge of it can both
improve a performance and answer a question of a convergence of the
algorithm in all possible situations, a question which is still open
(although the algorithm converged in all situations considered by
the author).

Secondly, although a numeric search for a rotation mentioned in the
previous paragraph takes a certain time, the most lengthy part is a
computation of a third moment of the lifted distribution. The author
made no essential attempt to optimize this part by, say,
implementing a kind of parallel computation. If there is a good way
to implement a parallel computation at this stage then the overall
performance of the algorithm becomes significantly better.

Despite these two shortcomings, the algorithm is rather fast if a
number of dimensions is not too large. For a five-dimensional data
of a one million points it works less than a half a minute on a
standard PC (a code is written in MATLAB). For large dimensionality
the algorithm in the form described above is inapplicable since it
requires at intermediate steps an addition of dimensions to data,
and their number goes as $N^2$. Then, however, one can implement a
different version of it, where one adds a single dimension at a time
and then rotates a distribution so as to minimize a norm of the
projection. Then again a dimension is added, etc. In such a version
there is no need in large amount of additional memory, but a
transformed data is no more a quadratic function of the initial one
but will rather be given by some high power functions, with the
power being uncontrollable. Similar ideas can be applied to
situations with exceptionally large higher moments, where the
algorithm discussed above might fail.

\section{Acknowledgements}

The author is grateful to O. Graubart for fruitful discussions along
the whole work.


\bibliographystyle{elsarticle-num}
\bibliography{<your-bib-database>}



\end{document}